\documentclass[apjl]{emulateapj}
 \usepackage{url}
\usepackage{amsmath}
\usepackage{xcolor} 
\usepackage{graphicx}
\usepackage[export]{adjustbox}
\usepackage{psfig}
\usepackage{multirow}
\usepackage[flushleft]{threeparttable}
\usepackage{amsmath}
\usepackage{multirow}
\usepackage[symbol*]{footmisc}
\DefineFNsymbolsTM{myfnsymbols}{
 \textasteriskcentered *
  \textdagger    \dagger
  \textsection   \mathsection
  \textbardbl    \|%
  \textparagraph \mathparagraph 
}%
\setfnsymbol{myfnsymbols}
 \begin{document}

\title{The first astrophysical result of HISAKI: a search for the EUV H\MakeLowercase{e} I lines in a massive cool core cluster at \MakeLowercase{z}=0.7}

\author{Yuanyuan Su\altaffilmark{$\ddagger$1,2}}
\author{Tomoki Kimura\altaffilmark{3}}
\author{Ralph P. Kraft\altaffilmark{2}}
\author{Paul E.\ J.\ Nulsen\altaffilmark{2}}
\author{Megan Gralla\altaffilmark{4}}
\author{William R.\ Forman\altaffilmark{2}}
\author{Go Murakami\altaffilmark{5}}
\author{Atsushi Yamazaki\altaffilmark{5}}
\author{Ichiro Yoshikawa\altaffilmark{6}}
\affil{$^1$Department of Physics and Astronomy, University of Kentucky, 505 Rose Street, Lexington, KY, 40506, USA}
\affil{$^2$Harvard-Smithsonian Center for Astrophysics, 60 Garden Street, Cambridge, MA 02138, USA}
\affil{$^3$Frontier Research Institute for Interdisciplinary Science, Tohoku University, Sendai, Japan}
\affil{$^4$Steward Observatory, University of Arizona, 933 N Cherry Avenue, Tucson, AZ 85721, USA}
\affil{$^5$ Institute of Space and Astronautical Science, Japan Aerospace Exploration Agency, Sagamihara, Japan}
\affil{$^6$Department of Complexity Science and Engineering, University of Tokyo, Kashiwa, Japan}
\altaffiltext{$\ddagger$}{Email: yuanyuan.su@cfa.harvard.edu}


\begin{abstract}
Molecular cold gas and star formation have been observed at centers of cool-core clusters, albeit at a level much smaller than expected from the classic cooling model. Feedback from the supermassive black hole is likely to have prevented hot gas from cooling. However, the exact cooling and heating processes are poorly understood. The missing key piece is the link between the hot gas ($10^7$\,K) and cold gas ($10^3$\,K). 
Using the extreme ultraviolet spectrometer onboard {\sl Hisaki}, we explore a distant galaxy cluster, RCS2 J232727.6-020437, one of the most massive cool-core clusters with a cooling rate of $400$\,M$_{\odot}$\,yr$^{-1}$. 
We aim to detect gas at intermediate temperatures ($3\times10^4$\,K) emitting He I$\alpha$ and He I$\beta$ at rest wavelengths of 58.4 nm and 53.7 nm, respectively. Our target resides at $z=0.6986$, for which these He I lines shift away from the absorption of the Galaxy. 
Our findings show that the amount of $10^{4-5}$\,K gas at the center of this cluster is smaller than expected if cooling there was uninhibited, which demonstrates that feedback both operates and is efficient for massive clusters at these epochs.

\end{abstract}

\section{\bf Introduction}

X-ray surface brightness increases rapidly towards the center of cool-core clusters (Forman \& Jones 1982). 
At the densities and temperatures observed in these regions, the X-ray emitting gas can cool catastrophically within a time much shorter than the Hubble time.
The classic cooling flow model predicts star formation rates of a few $100$\,M$_{\odot}$\,yr$^{-1}$ fueled by the condensation of the X-ray emitting hot gas. In contrast, the observed star formation rates are typically below $10$\,M$_{\odot}$\,yr$^{-1}$ (e.g., O'Dea et al.\ 2008). To reconcile the discrepancy, energy released by the active galactic nuclei (AGN) is invoked to compensate for the loss from the radiative cooling (McNamara \& Nulsen 2007). Cooling and heating are continuous processes. 
The entire feedback loop could be mapped out best if we knew the distribution of gas at all temperatures between the hottest ($\sim3\times10^7$\,K) and coolest ($\sim30$\,K) phases.
High-resolution X-ray spectroscopy of {\sl XMM-Newton}-RGS puts upper limits on the amount of $10^6$\,K gas via the observations of the Fe XVII line emission (e.g., Peterson et al.\ 2003). {\sl FUSE} UV observations of the O VI line have probed the bulk of gas at $10^{5-6}$\,K (Oegerle et al.\ 2001; Bregman et al.\ 2006). These studies suggest that gas at intermediate temperatures is deficient relative to the expectations of radiative cooling models. 

We present the first line emission observations of He I$\alpha$ at rest wavelengths of 58.43 nm and He I$\beta$ at 53.70 nm from a cool-core cluster, seeking for gas at $10^{4\--5}$\,K. {He I$\alpha$ and He I$\beta$ lines have long been predicted in atomic database (Dere et al. 1997; Landi et al. 2013; Ferland et al.\ 2017). The measurements of these neutral helium lines in intergalactic clouds using high redshift quasars can reveal the helium abundance in the early Universe, putting unique constraints on Big Bang nucleosynthesis (Reimers \& Vogel 1993; McQuinn \& Switzer 2010). He I$\alpha$ and He I$\beta$ lines have also been studied extensively in the solar plasma (Golding et al.\ 2017), i.e., providing strong diagnostics of motions in prominences (Labrosse et al.\ 2007). 
He I$\alpha$ and He I$\beta$, both peaked at a temperature of $3.16\times10^4$\,K, are ideal tracers of gas at intermediate temperatures.} In the intracluster medium (ICM), Helium is much more abundant than Iron and Oxygen.    
Despite the rich physics they can probe, these neutral helium lines have never been studied in clusters of galaxies. 
For nearby clusters, the wavelengths of He I$\alpha$ and He I$\beta$ lie short-ward of the Lyman edge. Galactic absorption would reduce the line flux by 3--4 orders of magnitude, making them effectively undetectable.  
Our target RCS2 J232727.6-020437 (hereafter RCS2327) is a galaxy cluster residing at $z=0.6986$, for which the wavelengths of He I$\alpha$ and He I$\beta$ have shifted to 99.25 nm and 91.21 nm, respectively, long-ward of the Lyman edge.

{\sl Hisaki} is a Spectroscopic Planet Observatory operated by JAXA. 
It was launched in 2013 as the first mission of the Small Scientific Satellite Project (Yoshioka et al.\ 2013).  
Its EUV spectrometer produces two-dimensional spectral image over the wavelength of 55--145 nm (Yoshikawa et al.\ 2014; Kimura et al.\ 2019). {{\sl Hisaki} observations are limited to targets with latitudes within $\pm10^{\circ}$ in the Ecliptic Coordinate. 
Over the past few years, Sunyaev–Zel’dovich (SZ; Sunyaev \& Zeldovich 1972) surveys  (i.e., Planck, SPT, ACT) have discovered a rapidly growing number of distant clusters. We searched through galaxy clusters in the SZ Cluster Database\footnote{\url{http://szcluster-db.ias.u-psud.fr/}} and found four clusters at $z\gtrsim0.7$ that are visible to {\sl Hisaki}. Among them, RCS2327 is the only cool core cluster.}
RCS2327 was first discovered in the Second Red-Sequence Cluster Survey (RCS2; Gilbank et al.\ 2011). It is one of the most relaxed and spherically-symmetric clusters. Sharon et al.\ (2015) have measured its total mass in every known way: hydrostatic X-ray measurement, weak-lensing, strong-lensing, SZ effect, dynamics of member galaxies, and galaxy richness. The authors found that all these methods give consistent results that RCS2327 is the most massive galaxy cluster at this redshift with an enclosed mass of $3\times10^{15}$\,M$_{\odot}$ within its virial radius ($R_{200}=2.1$\,Mpc). 

We present joint {\sl Hisaki} EUV and {\sl Chandra} X-ray observations of RCS2 J232727.6-020437. We adopt the cosmological parameters $H_0=70$ km s$^{-1}$ Mpc$^{-1}$, $\Omega_{\Lambda}=0.73$ and $\Omega_m=0.27$. This cosmology gives a linear scale of $1^{\prime\prime} = 7.24$ kpc at the cluster redshift of $z=0.6986$. Uncertainties reported are quoted at a confidence level of 68\% throughout this work.

\section{\bf observations and data reductions}

\subsection{Hisaki}

\begin{figure*}
\vspace{-1.3cm}
   \centering
           \includegraphics[width=0.95\textwidth]{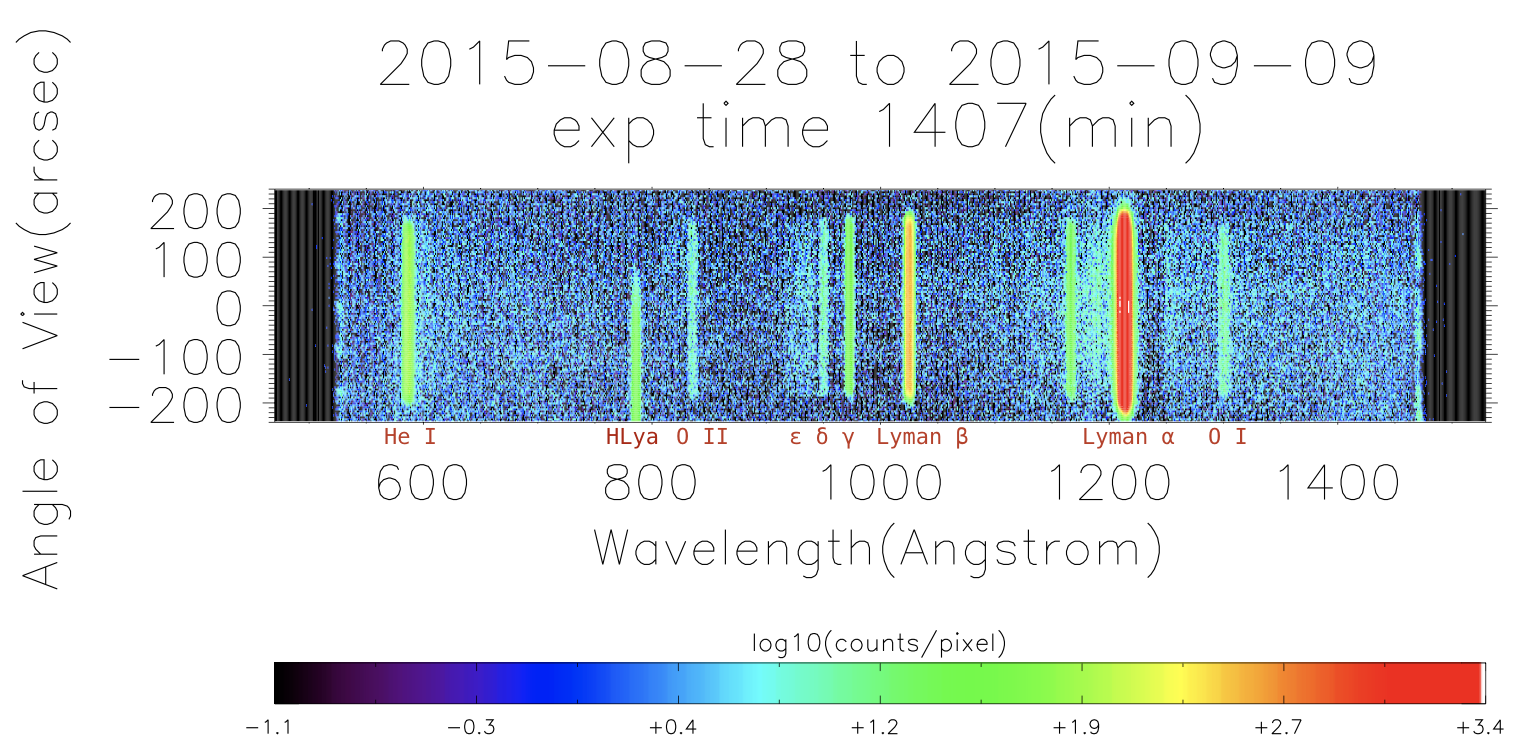}
\figcaption{\label{fig:slit} The {\sl Hisaki} EUV imaging-spectral distribution of RCS2327. 
Our target is centered at a viewing angle of $110\arcsec$ on the detector. No obvious emission is detected other than geocoronal lines.}
\end{figure*}

\begin{figure*}
   \centering
           \includegraphics[width=0.95\textwidth]{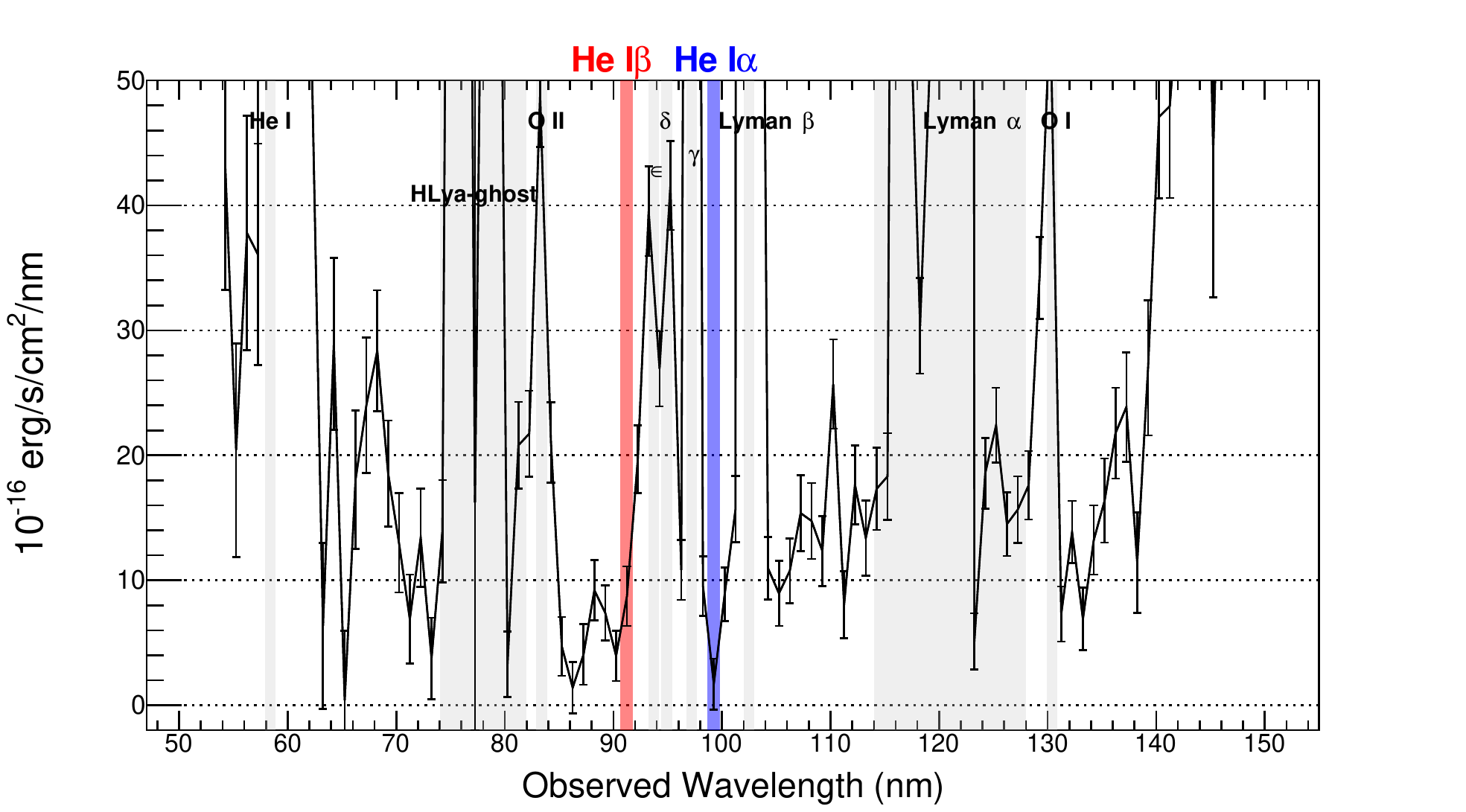}
\figcaption{\label{fig:spec1} The {\sl Hisaki} spectrum for the viewing angle range $(100{\arcsec}, 120{\arcsec})$ over the wavelength range 55-150 nm with instrumental background subtracted. Geocoronal emission lines and the HLy$\alpha$ ghost line are marked in gray. The He I$\alpha$ and He I$\beta$ lines of RCS2327 are shifted to 99.25 nm and 91.21 nm, respectively, at $z=0.7$.}
\end{figure*}

{\sl Hisaki} made three observations of RCS2327 in August and September 2015. All the observations were taken during orbital night with a total exposure time of 84.4\,ksec. A slit width of 60$^{\prime\prime}$ was used, which gives a spectral resolution of $\lesssim$\,10\,$\mathring{A}$ and a spatial resolution of $\lesssim\,17\arcsec$. Only Level 2 data\footnote{Level 2 data for the three observations are: 
exeuv.rcs2\_j23\_h-15v-89\_n.mod.01.20150828.lv.02.vr.00.fits, exeuv.rcs2\_j23\_h-15v-89\_n.mod.01.20150906.lv.02.vr.00.fits, and exeuv.rcs2\_j23\_h-15v-89\_n.mod.01.20150909.lv.02.vr.00.fits. The data are archived in the Data Archives and Transmission System (DARTS) JAXA ({\url http://www.darts.isas.jaxa.jp/stp/hisaki/}).} were included in our analysis. Data reduction was performed with the IDL Astronomy User's Library and NAIF SPICE toolkit. 
A spectral image integrated over all observations is shown in Figure~\ref{fig:slit}. 
The {\sl Hisaki} detector extends over the full angular range $(-1000\arcsec, 600\arcsec)$, but only the region in the range $(-180\arcsec, 180\arcsec$) is exposed to the sky. We extract instrumental background from the range $(200\arcsec, 300\arcsec)$. 
Our target is centered at $110\arcsec$. A spectrum of RCS2327 over ($100\arcsec, 120\arcsec$) with instrumental background subtracted is shown in Figure~\ref{fig:spec1}. No obvious emissions are detected other than geocoronal emission lines (Kuwabara et al.\ 2017).

\begin{figure}
   \centering
    \includegraphics[width=0.51\textwidth]{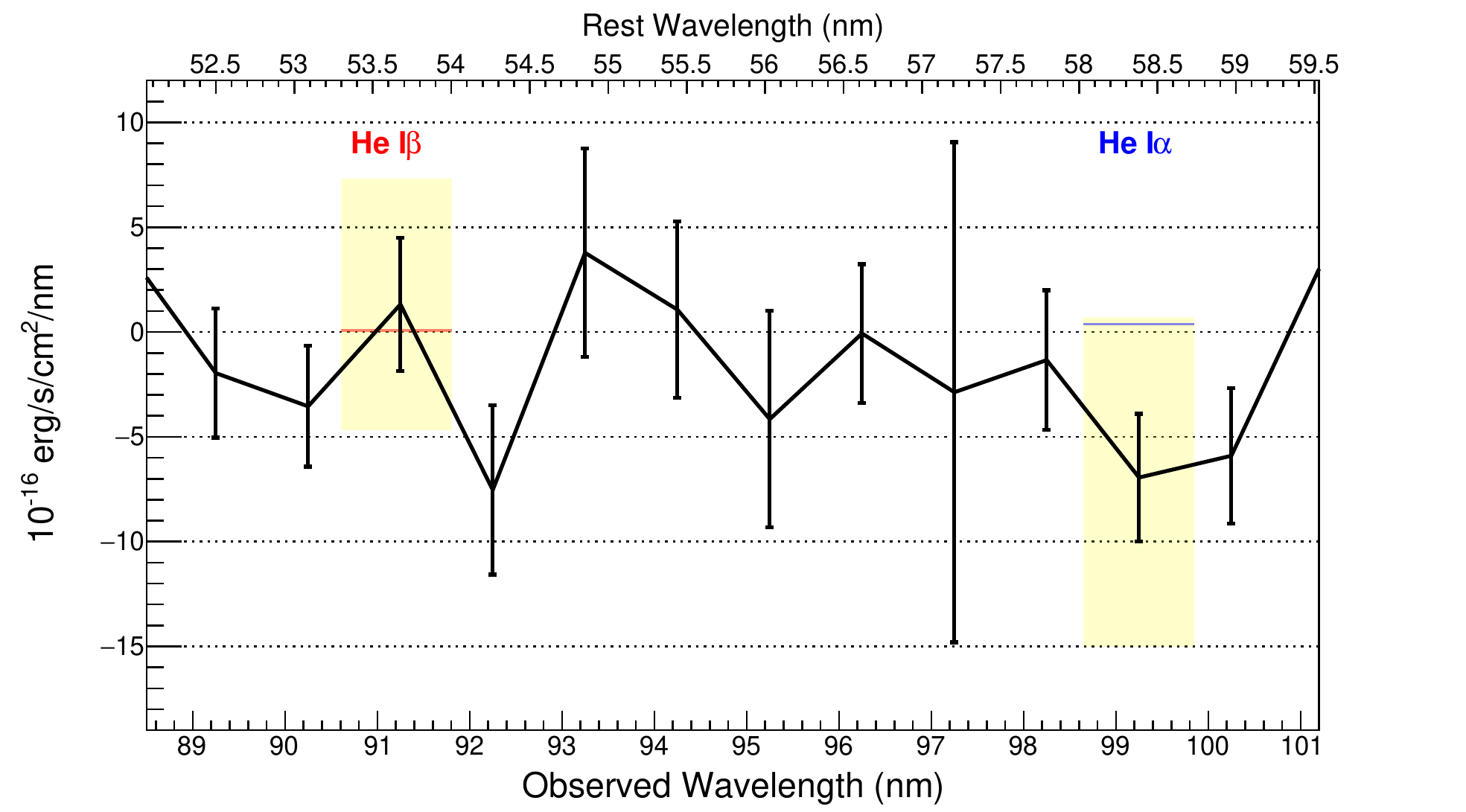}
\figcaption{\label{fig:spec2} The {\sl Hisaki} spectrum of our target lines (He I$\alpha$ and He I$\beta$ are at 58.4 nm and 53.7 nm  in the rest frame, respectively). The spectrum is produced by subtracting the $-120{\arcsec}\sim-100{\arcsec}$ spectrum from the $100{\arcsec}\sim120{\arcsec}$ spectrum. X-axis error bar represents the spectral resolution of 1 nm. Red and blue lines mark the line fluxes expected from the steady cooling flow model. Black error bars indicate the statistical uncertainties alone. Yellow regions represent the sum of the statistical and systematic uncertainties added in quadrature.} 
\end{figure}

The sky background is not uniformly distributed over the field of view of {\sl Hisaki}.
This is critical for faint sources like our target.  
We calibrate the spatial variation of the local background based on the distribution of geocoronal emission lines. 
The HLy$\alpha$ line around 80\,nm is a ghost line and should be ignored. The spatial distribution of the Lyman $\alpha$ line has been severely affected by the dumbbell
slit mode for Jupiter monitoring. Therefore we only consider geocoronal emission lines of Lyman $\beta$ (102.5 nm), O II (83.4 nm), O I (130.4 nm), and He I (58.4 nm).  
We obtain the total flux of these four lines (with a spectral bin size of 1 nm) as a function of viewing angle: $f_{\rm geo}(\theta)$. We normalize the flux of He I$\alpha$ and He I$\beta$ as a function of view angle by $f_{\rm geo}(\theta)$. We extract the signal spectrum from $100\arcsec\sim120\arcsec$, corresponding to a region centered on the cluster X-ray peak with a radius of 74\,kpc. We extract a local background from $-120\arcsec\sim-100\arcsec$, which consists of both astrophysical and instrumental backgrounds. The resulting background subtracted spectrum zoomed in on  He I$\alpha$ and He I$\beta$ is shown in Figure~\ref{fig:spec2}.

\subsection{Chandra}

\begin{figure}
   \centering
           \includegraphics[width=0.5\textwidth]{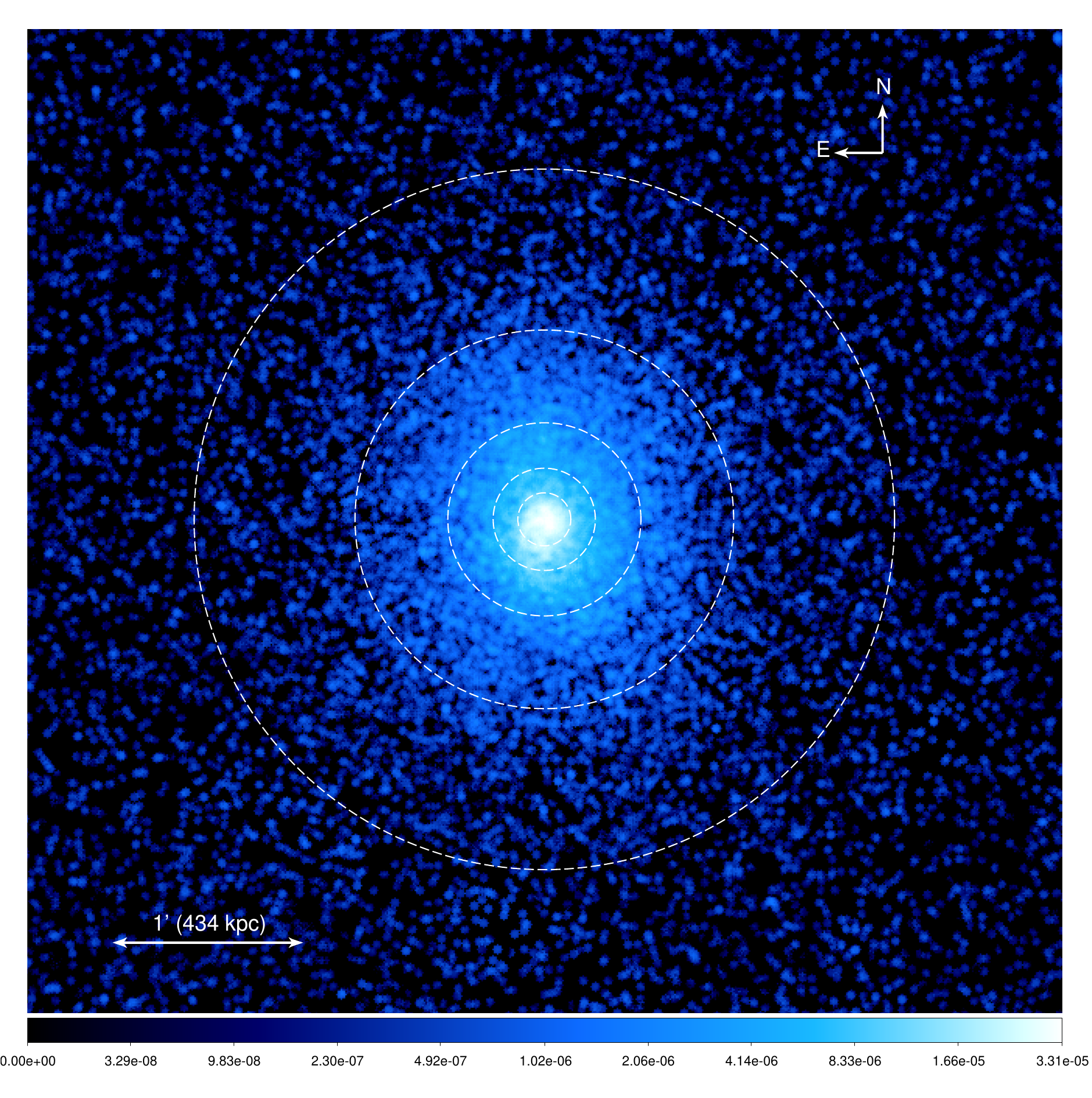}
\figcaption{\label{fig:chandra} {\sl Chandra} X-ray image of RCS2327 in the 0.5--2.0 keV energy band in units of photon\,cm$^{-2}$\,s$^{-1}$. Its X-ray morphology is symmetric with no significant substructures detected in the ICM, suggesting that RCS2327 is a relaxed cluster. Regions used for spectrum analysis are marked in white annuli.}
\end{figure}

We use the {\sl Chandra} X-ray Observatory to probe the ICM properties of RCS2327. The superb spatial resolution of {\sl Chandra} makes it ideal to study this cluster at high redshifts. 
Our analysis includes all three {\sl Chandra} observations of RCS2327: one 25\,ksec ACIS-S pointing (ObsID: 7355) taken in August 2007 and two $\sim$75\,ksec ACIS-I pointings (ObsID: 14025 and 14361) obtained in 2011. CIAO 4.8 and CALDB 4.6.9 were used for the data reduction.
{\tt chandra\_repro} was used to reprocess all the observations from level 1 events. Background flares beyond 3$\sigma$ were filtered using the light curve filtering script {\tt lc\_clean}. We obtain a total of 166 ksec cleaned exposure time. Point sources were detected in a 0.3--7.0 keV image with {\tt wavdetect}. 
{A mosaic image of RCS2327 is shown in Figure~\ref{fig:chandra}; its X-ray morphology is symmetric with no significant substructures detected in the ICM, suggesting that RCS2327 is a relaxed cluster.  
We extract spectra from 5 consecutive annuli centered on the cluster center as shown in Figure~\ref{fig:chandra}, so that each annulus contains at least 5000 net counts.} The innermost region has a radius of 61\,kpc (8.4 arcsec). 
Tailored blank sky spectra generated with CIAO tool {\tt blanksky} were applied as the background spectra. All spectra were grouped to have at least one count per energy bin. Spectral fitting was performed with {\sl XSPEC\,12.9.1} using the C-statistic.
Deprojection was performed by fitting the spectra with the mixed model: ${\tt projct}\times({\tt phabs}\times{\tt vapec})$.  
The solar abundance standard of Asplund at al.\ (2006) and a Galactic 
hydrogen column of $N_{\rm H}=4.8\times10^{20}$ cm$^{-2}$ (Kalberla et al.\ 2005) are adopted.

\section{\bf Results}

\subsection{He I line emission}

\begin{figure}
   \centering
       \includegraphics[width=0.5\textwidth]{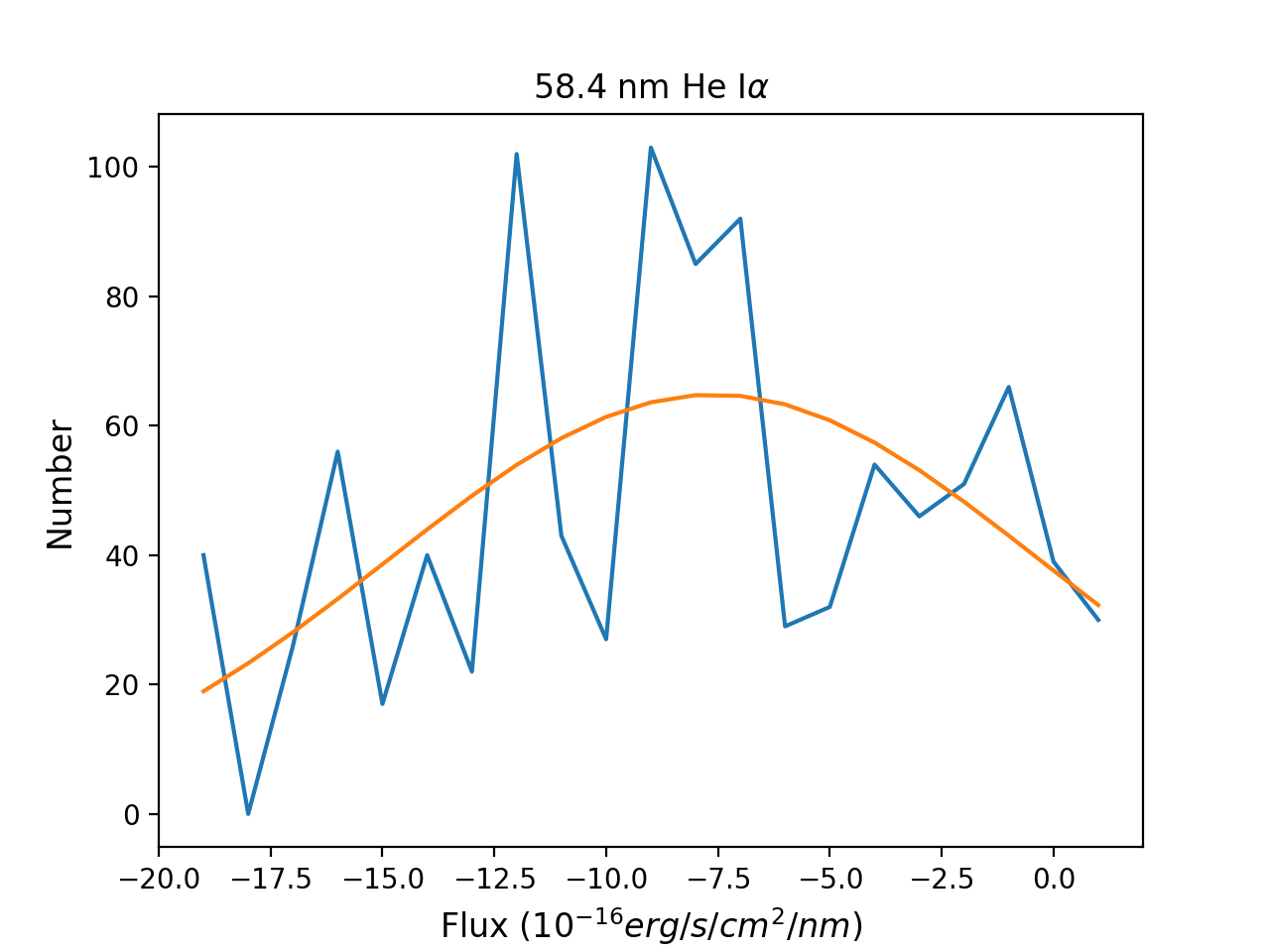}
       \includegraphics[width=0.5\textwidth]{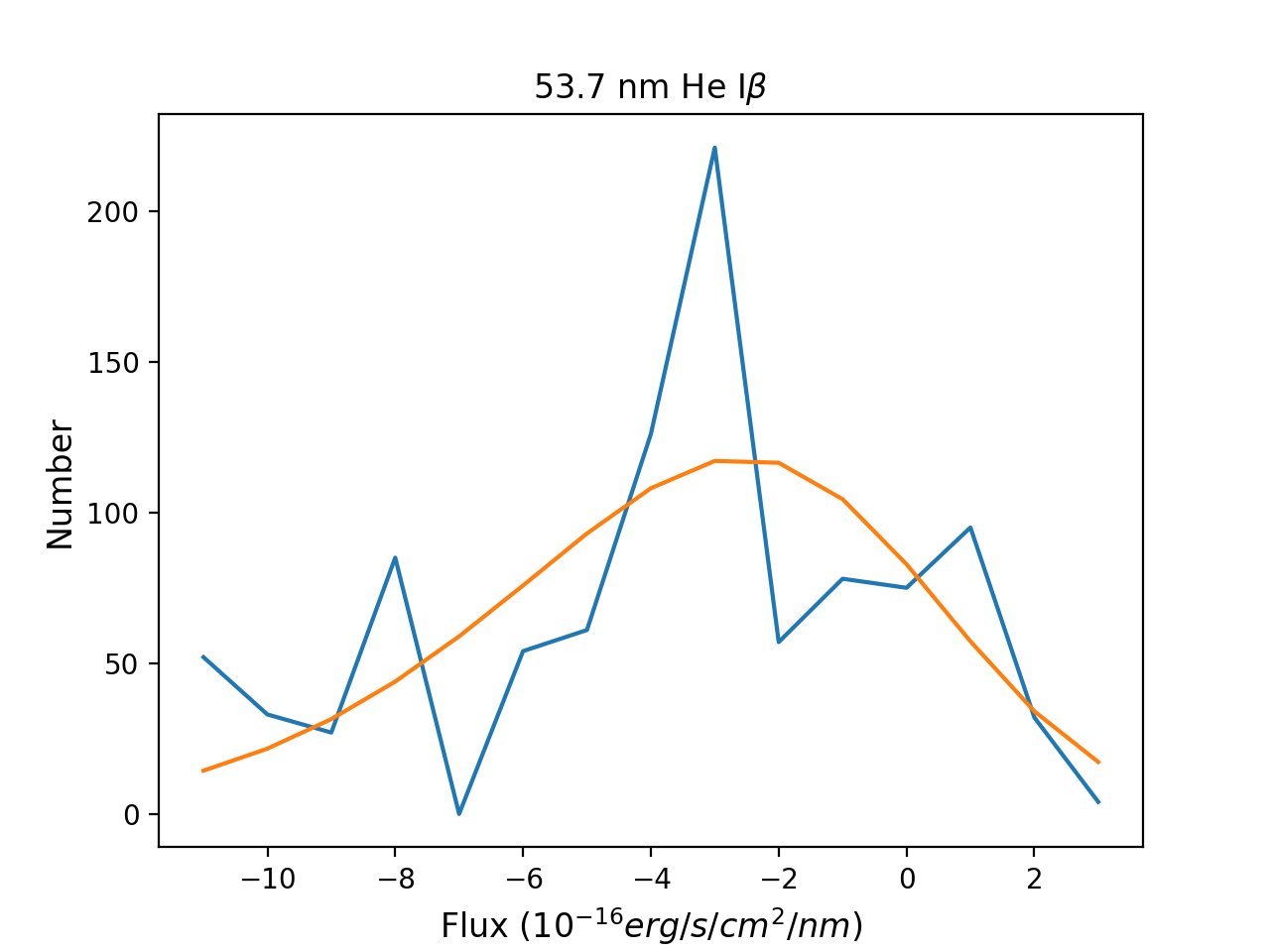}
       
\figcaption{\label{fig:sys} The distribution of the fluxes of He I$\alpha$ (top) and He I$\beta$ (bottom) of $100{\arcsec}\sim120{\arcsec}$ measured by using 40 different $20\arcsec$ wide regions as local background. {The histograms were fit to skew normal distributions}. The $1\sigma$ uncertainties are 7.0 and 5.1 $\times10^{-16}$ erg\,s$^{-1}$\,cm$^{-2}$ for the top and bottom figures respectively.}
\end{figure}

We present the {\sl Hisaki} spectrum of the central $r<74$\,kpc of RCS2327 in Figure~\ref{fig:spec1}, corresponding to ($100\arcsec\sim120\arcsec$) on the detector. A zoomed-in view of He I$\alpha$ and He I$\beta$ is presented in Figure~\ref{fig:spec2} with local background ($-120\arcsec\sim-100\arcsec$) subtracted. Their fluxes are consistent with a non-detection. 
{The absence of He I$\alpha$ and He I$\beta$ line emissions is unexpected in a steady cooling flow scenario in which X-ray emitting gas cools at an estimated cooling rate of hundreds of solar mass per year. However, their upper limits still allow some level of cooling that we will discuss in \S4. 

The systematic uncertainties are dominated by the spatial variation of the instrumental backgrounds and the airglow foregrounds.}
To access the systematic uncertainties, we repeat the analysis by using each of the following 40 regions as the local background: $-160\arcsec\sim-140\arcsec$, $-155\arcsec\sim-135\arcsec$, $-150\arcsec\sim-130\arcsec$,  ... $30\arcsec\sim50\arcsec$, $35\arcsec\sim55\arcsec$, $40\arcsec\sim60\arcsec$. The resulting flux of He I$\alpha$ and He I$\beta$ are shown in Figure~\ref{fig:sys}. 
We fit their flux distributions to a skew-normal model.
We take the standard deviation of the flux distribution as the systematic uncertainty. 
The measured fluxes of He I$\alpha$ and He I$\beta$ are $-6.9\pm 3.1 ({\rm sta})\pm7.0 ({\rm sys})\times10^{-16}$ erg\,s$^{-1}$\,cm$^{-2}$ and $1.3\pm 3.2 ({\rm sta})\pm5.1 ({\rm sys})\times10^{-16}$ erg\,s$^{-1}$\,cm$^{-2}$, respectively. 

\subsection{X-ray emitting intracluster medium}

\begin{figure}
   \centering
           \includegraphics[width=0.5\textwidth]{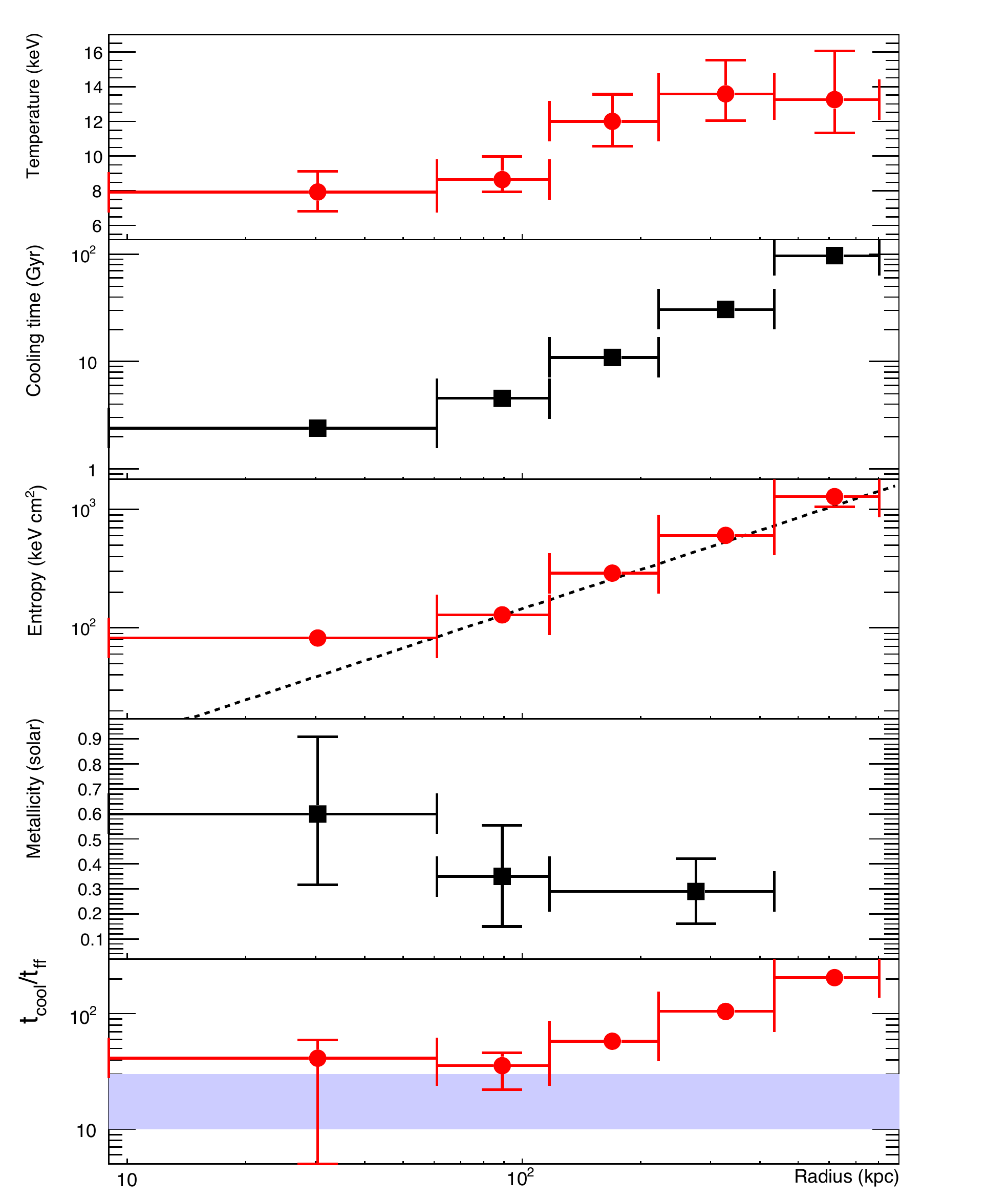}
\figcaption{\label{fig:profile} Deprojected radial profiles of temperature, cooling time, entropy, metallicity, and $t_{\rm cool}/t_{\rm ff}$, centered on RCS2327 obtained with {\sl Chandra}. Black dashed line represents the power law model of the entropy profile with a best-fit slope of $0.94\pm0.07$. The blue shaded region marks the range of the threshold $t_{\rm cool}/t_{\rm ff}$ to induce thermal instability.}
\end{figure}
We performed a deprojected analysis of the hot gas in RCS2327 using {\sl Chandra} observations as shown in Figure~\ref{fig:profile}. The best-fit temperature is 8\,keV at the cluster center and increases to 14\,keV at 300\,kpc ($\sim0.15$\,$R_{\rm vir}$). 
We calculate the gas cooling time, $t_{\rm cool}$, using
 \begin{equation}\label{eq:4}
 t_{\rm cool}=\frac{3P}{2n_en_H\Lambda(T,Z)},
\end{equation}
where $P=1.8n_ekT$ is the pressure and $\Lambda(T,Z)$ is the cooling function determined by the plasma temperature and metallicity. 
The three-dimensional entropy profile is derived with the temperature and density profiles, $K=kT\,{n_e}^{-2/3}$.  
Numerical simulations and observations of high redshift clusters suggest that most massive clusters have assembled around $z=2$ (Wang et al.\ 2016; Chiang et al.\ 2013). At the current redshift of RCS2327, 4\,Gyr has passed since its formation. Only the innermost bin has a cooling time shorter than 4\,Gyr, defining the regime of the cool core.   
A central metallicity excess, probably produced by supernova explosions in the brightest cluster galaxy, is also typical for relaxed cool-core clusters. 
The innermost bin of our analysis has a metallicity of 0.6\,$Z_{\odot}$ while the outer region displays a more uniformly distributed metallicity of 0.3\,$Z_{\odot}$, {although the existing data is not deep enough to indicate a significant central metallicity excess.}   
We calculate the cooling rate via
\begin{equation}\label{eq:15}
\dot{M}_{\rm cool}=\frac{2}{5}\frac{\mu m_pL_X}{kT_X},
\end{equation}
where $L_{\rm X}$ is the bolometric X-ray luminosity and $\mu=0.62$ the mean molecular weight. 
Its cooling rate, integrated over the cool core, is $\dot{M}_{\rm cool}=411$\,M$_{\odot}/$yr, indicating a strong cooling flow.

\section{\bf Discussion}
\begin{figure}
   \centering
               \includegraphics[width=0.5\textwidth]{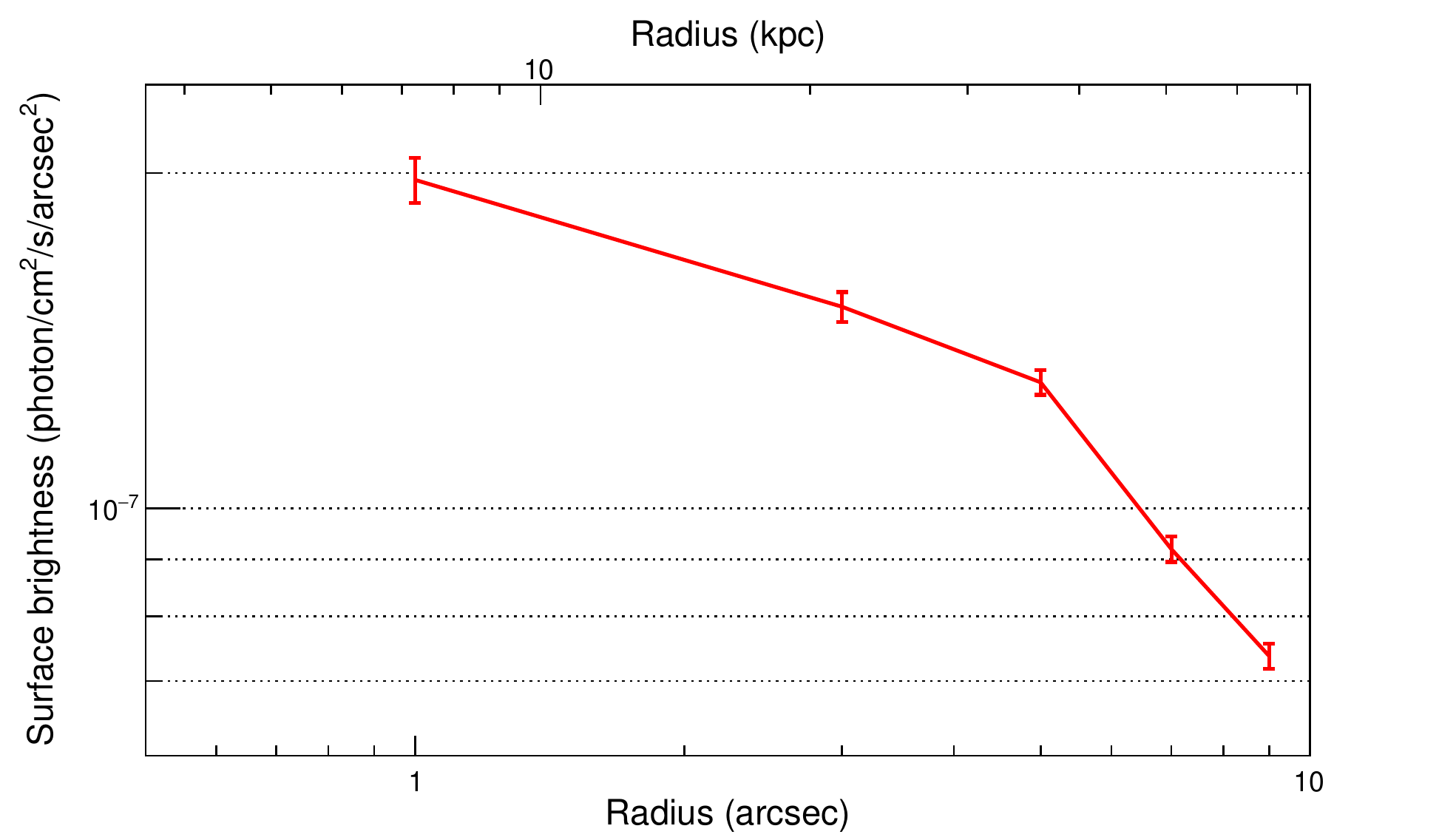}
\figcaption{\label{fig:sur} {\sl Chandra} surface brightness profile in the 0.5--7.0 keV energy band of the innermost bin in the spectral analysis. The surface brightness profile rises steeply towards the cluster center. 
The actual cooling at the cluster center is likely to be more vigorous than what we could infer spectroscopically with existing {\sl Chandra} data.}
\end{figure}

We have performed the first EUV observation of a galaxy cluster using the {\sl Hisaki} Planetary Observatory. This is {\sl Hisaki}'s first observation of an astrophysical object. We aim to probe gas at $10^{4-5}$\,K via measuring the flux of He I lines.
Our target RCS2327 resides at $z=0.7$. The wavelengths of He I$\alpha$ and He I$\beta$ are redshifted to 99.25 nm and 91.21 nm, respectively. These wavelengths are in the range of {\sl Hisaki} and long-ward of the Lyman edge of the Galaxy. 
RCS2327 is one of the most massive and relaxed clusters in the Universe. 
The innermost region in our {\sl Chandra}
analysis has a radius of 61 kpc. Its deprojected entropy index is
$80\rm\ keV\ cm^{2}$ and its cooling time is 2.4\,Gyr. Typical nearby cool-core clusters have an entropy as low as 10\,keV\,cm$^2$ and a cooling time of $<1$\,Gyr within $r<10$\,kpc (Panagoulia et al.\ 2014).  
We fit the entropy profile to a power law model and obtain a best-fit slope of $0.94\pm0.07$. Extrapolating this best-fit profile to the inner 10\,kpc, its entropy would reach below 20\,keV\,cm$^2$. 
The surface brightness within the innermost bin keeps rising towards the cluster center as shown in Figure~\ref{fig:sur}. 
The actual cooling at the cluster center is likely to be more vigorous than what we could infer spectroscopically with existing {\sl Chandra} data.

\subsection{Gas mass at intermediate temperatures}

The flux of He I lines allows us to probe its cluster gas at $\sim3\times10^4$\,K. The upper limit of the He I$\beta$ line is substantially higher than that of the He I$\alpha$ line (\S3.1). Therefore, we use the measurement of the He I$\beta$ line to put upper limits on the mass and volume of the gas at intermediate temperatures at the center of RCS2327. 
We calculate $\Lambda_{\rm line}n_en_{\rm H}$, the energy loss rate per unit volume, where $\Lambda_{\rm line}$ is the contribution to the cooling function from each line and the ratio of $n_e$ to $n_{\rm H}$ is 1.2.
We add the statistical and systematic uncertainties in quadrature to obtain the $1\sigma$ upper limit of $7.8\times10^{-16}$\,erg\,s$^{-1}$\,cm$^{-2}$ for He I$\beta$, 
which requires the volume of the $\sim3\times10^4$\,K gas to be $<1.2\times10^{6}$ pc$^3$.
We assume that the warm gas and the X-ray emitting hot gas are in pressure equilibrium. 
We obtain that the $\sim3\times10^4$\,K gas mass should be no more than $2.9\times10^6\,$M$_{\odot}$. 
As we discuss below, gas at intermediate temperatures provides unique constraints on the interaction between the hot gas and the cold gas, particularly the direction of the energy flow. {We demonstrate the implications of the lack of warm gas in RCS2327 via making analogies with the brightest galaxy cluster in X-ray, Perseus and the nearest galaxy cluster, Virgo.
 
\subsubsection{Cooling}

Gas at intermediate temperatures can be produced via the cooling of the X-ray emitting hot gas. The nearby cluster, Perseus, is the brightest galaxy cluster in X-ray (Forman et al.\ 1972). Its cool core has been observed extensively over the entire electromagnetic spectrum. 
In particular, extended filamentary nebula of $10^4$\,K has been observed surrounding its BCG NGC~1275 (Conselice \& Gallagher 1999; Gendron-Marsolais et al.\ 2018). Up to 5$\times10^{10}$M$_{\odot}$ of cold molecular gas has been found to be associated with these H$_{\alpha}$ filaments (Salom$\rm \acute{e}$ et al.\ 2006). The stellar body of NGC~1275 contains very blue star clusters which is unusual for an early-type galaxy (Conselice et al.\ 2001). The cool core of Perseus clearly contain multi-phase materials, which is not rare for nearby cool core clusters (McDonald et al.\ 2010). The rich phenomena in NGC~1275 may be explained by the residual cooling coupled with the mechanical feedback of its AGN (Salom$\rm \acute{e}$ et al.\ 2006). 
We note that the cool core of Perseus cluster has a X-ray-derived mass deposition rate of $\dot{M}_{\rm cool}=300$\,M$_{\odot}/$yr (Fabian et al.\ 2000) which is comparable to that of RCS2327 of 400\,M$_{\odot}/$yr.
If Perseus were like RCS2327 with little warm gas at $3\times10^4$\,K, it would undermine the role of the cooling of X-ray emitting hot gas; instead, recent mergers with gas-rich galaxies may be invoked to interpret the multi-phase cluster gas (Conselice et al.\ 2001).}

\begin{figure}
   \centering
           \includegraphics[width=0.5\textwidth]{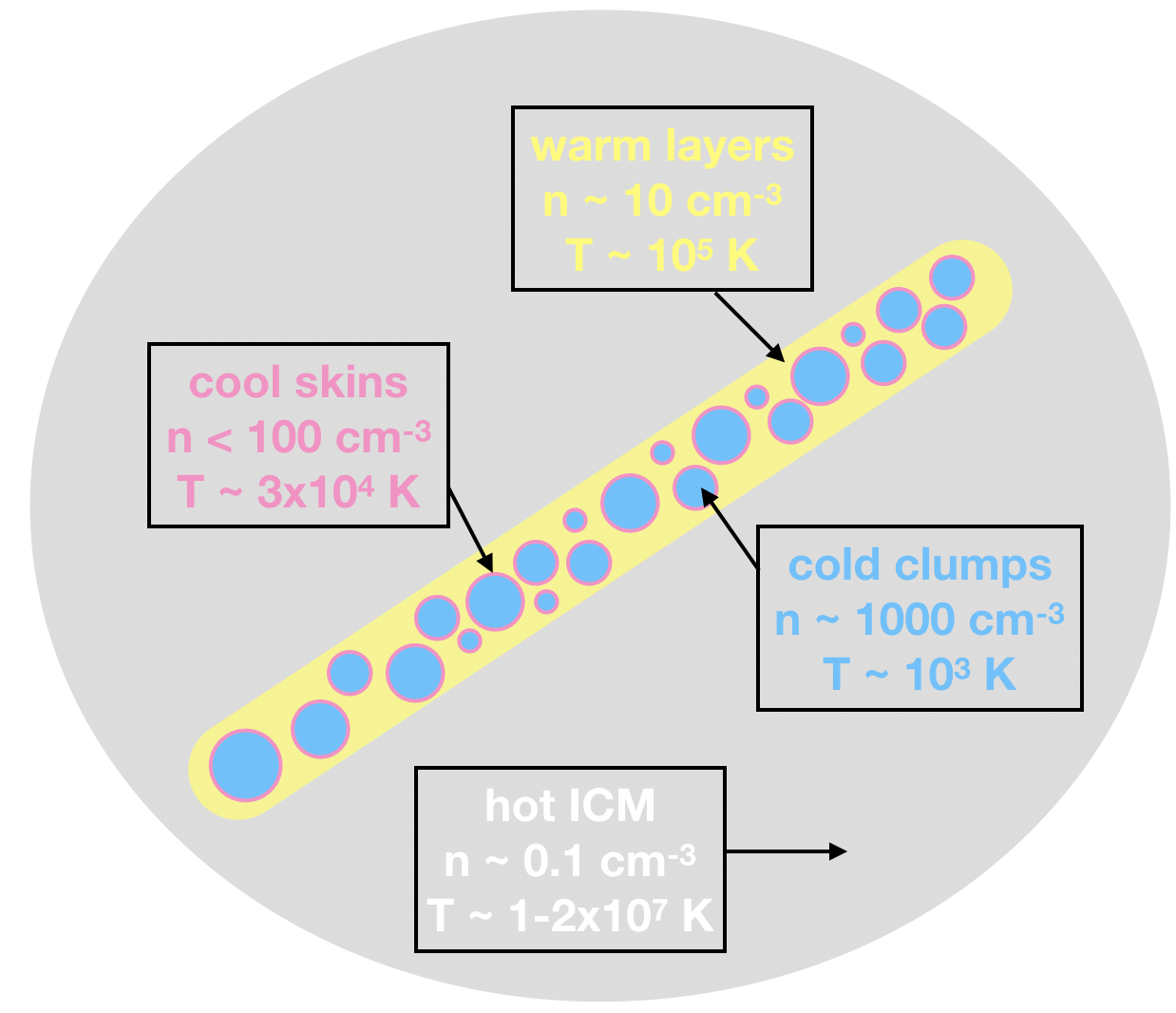}
\figcaption{\label{fig:sketch} A sketch demonstrates the possible structure of the multiphase filaments found at centers of cool core clusters as implied by our observations of RCS2327 (The sketch is based on Anderson \& Sunyaev 2018; Werner et al.\ 2019).}
\end{figure}

\subsubsection{Conduction}

We estimate that the volume of the $\sim3\times10^4$\,K gas is no more than $1.2\times10^{6}$ pc$^3$ in RCS2327.
If such warm gas is distributed throughout the cool core, it would have an extremely small volume filling factor ($f<10^{-6}$). The upper limit we obtain allows it to exclusively occupy individual filaments (i.e., with a radius of $r_c<100$\,pc). Any warm gas in RCS2327 is likely to be confined in filament structures.   
Using HST/COS observations, Anderson \& Sunyaev (2018) studied FUV emission along the line-of-sight to a filament in M87, the BCG of the Virgo cluster. The authors infer that warm gas ($T\sim10^5$\,K, $n\sim10$\,cm$^{-3}$) forms a very thin layer around cold blobs ($T\sim10^3$\,K, $n\sim1000$\,cm$^{-3}$), together imbedded in the hot ICM ($T\sim10^7$\,K, $n\sim0.1$\,cm$^{-3}$). 
If the constraints we put on the $3\times10^4$\,K gas ($n\lesssim100$\,cm$^{-3}$) can be applied to the filament in M87, the $3\times10^4$\,K gas may form a third layer between the $T\sim10^5$\,K gas and the cold clumps. A possible structure of the multi-phase filament is demonstrated in Figure~\ref{fig:sketch}.
Cold gas evaporating through conductive heating would be a natural explanation of these multi-phase gases that are in pressure equilibrium.  
The lack of warm gas would otherwise suggest quenched conduction.

\subsubsection{Mixing}
Cold gas could flow through the ambient hot gas to reach the supermassive black hole. A highly turbulent ICM would promote Kelvin--Helmholtz instability to develop at the interface to form a layer of mixed gas (Begelman \& Fabian 1990; Su et al.\ 2017a,b). 
The expected gas temperature resulting from the mixing between the $10^3$\,K cold gas and the $10^7$\,K hot gas should be approximately $\sqrt{T_cT_h}=10^5$\,K (Fabian et al.\ 2001). If the turbulent mixing is effective, the $10^5$\,K gas may overshadow gas at higher or lower temperatures (Fabian et al.\ 2001; Oegerle et al.\ 2001). This scenario is consistent with previous observations and our results: {\sl FUSE} has speculatively detected gas near $10^5$\,K in cool-core clusters: O VI emission at $3\times10^5$\,K and C III emission at $7\times10^4$\,K in Abell~2597 (Oegerle et al.\ 2001) and possibly also in Abell~1795 (Bregman et al.\ 2006). The {\sl XMM-Newton} RGS spectra of Abell~1835 indicates a lack of Fe XVII emission at $4\times10^6$\,K (Peterson et al.\ 2003) and our study reveals a non-detection of He I lines at $3\times10^4$\,K in RCS2327.
{\it Hitomi} observations of the Perseus Cluster, however, indicate a rather quiescent ICM with a turbulent velocity of $\sim$150\,km\,s$^{-1}$ (Hitomi Collaboration et al.\ 2016). Yet the existing UV-EUV observations and the {\sl Hitomi} observations are too few and none of these observations are performed for the same cluster.

\subsection{Steady cooling flow}

We expect emission lines to be produced by the cooling plasma. 
We calculate the expected line emission luminosities of He I$\alpha$ and He I$\beta$ following 
Edgar \& Chevalier (1986),
\begin{equation}
L_{\rm line}=k\frac{\dot{M}_{\rm cool}}{\mu m_{\rm H}}A_{\rm Z}\int^{T_{\rm max}}_{T_{\rm min}}\left(\frac{3}{2}+s\right)\chi\frac{f_i j_{\rm line}}{\Lambda}\frac{hc}{\lambda_{\rm line}}dT,
\end{equation}
where $T_{\rm min}=10^4$\,K and $T_{\rm max}=10^6$\,K, $\dot{M}_{\rm cool}$ is the cooling rate (see \S3), $A_{\rm Z}$ the abundance of He relative to H, $\chi$ the number of particles per H atom. 
{The cooling process may be in an intermediate state between 
isobaric ($s=1$) and isochoric ($s=0$),
therefore we choose $s=0.5$.}
The ionization fraction, $f_i (T)$, and the line emissivity coefficient, $j_{\rm line}$, are derived from {\tt AtomDB}.
Substituting the cool core parameters of RCS2327, we obtain an expected flux of $3.67\times10^{-17}$ erg\,s$^{-1}$\,cm$^{-2}$ and $3.85\times10^{-18}$ erg\,s$^{-1}$\,cm$^{-2}$ for He I$\alpha$ and He I$\beta$, respectively. 
The expected flux of He I$\alpha$ is 10 times that of He I$\beta$. 
The line emission of He I$\alpha$ is not detected by {\sl Hisaki} with a flux of $-6.9\pm 3.1 ({\rm sta})\times10^{-16}$ erg\,s$^{-1}$\,cm$^{-2}$, suggesting a lack of steady cooling flow in RCS2327. 
{However, after adding the statistical and systematic uncertainties in quadrature, we obtain a 1$\sigma$ upper limit of $7.5\times10^{-17}$ erg\,s$^{-1}$\,cm$^{-2}$ for He I$\alpha$. Therefore, we can not firmly rule out the cooling flow in a steady state in RCS2327.}
The uncertainty of the He I$\beta$ flux is too significant to provide meaningful constraints on the cooling process.

\subsection{Thermally unstable cooling}

Reservoirs of cold gas detected at centers of cool-core clusters cast new light on the fueling of supermassive blackholes (Edge 2001; Russell et al. 2014, 2016, 2017; Vantyghem et al. 2016; McNamara et al.\ 2016). The prevailing AGN feedback model suggests that thermal instability can happen if the ratio of the cooling time $t_{\rm cool}$ and the free-fall time $t_{\rm ff}(=\sqrt{2r/g})$ drops below a threshold value (McCourt et al.\ 2012). Cold gas, condensed from the hot gas, could be accreted onto the super massive black hole to trigger AGN feedback (Gaspari et al.\ 2013). 
Massive galaxy clusters that do contain cold gas seem to have a $t_{\rm cool}/t_{\rm ff}$ ratio $\lesssim20$ (Hogan et al.\ 2017). 
We present the $t_{\rm cool}/t_{\rm ff}$ profile of RCS2327 in Figure~\ref{fig:profile}.
The total mass profile taken from Sharon et al.\ (2015) was used to compute the gravitational constant g.  
The measured values of $t_{\rm cool}/t_{\rm ff}$ in the central region of RCS2327 are consistent with values below 20.
The absence of gas at intermediate temperatures implies that there may be a delay between the drop of $t_{\rm cool}/t_{\rm ff}$ and the onset of the thermal instability.
Then again, the uncertainty of its $t_{\rm cool}/t_{\rm ff}$ is too significant and its cold gas content is unknown; RCS2327 may not satisfy the condition to induce thermal instabilities.

\section {\bf Summary}

This work presents the first measurement of He I lines at the center of a cool-core galaxy cluster, probing cluster gas at $10^{4-5}$\,K.
Our target, RCS2 J232727.6-020437, is a massive cluster at $z=0.7$, 
placing its redshifted He I lines long-ward of the Lyman edge at $z=0$.  
Currently, there are no active instruments intended for extragalactic EUV studies. 
The EUV observations used in this study were performed with the {\sl Hisaki} Planet Observatory. We put upper limits on the He I$\alpha$ and He I$\beta$ line emission.  
The mass of the $3\times10^4$\,K gas is no more than $2.9\times10^6\,$M$_{\odot}$, occupying a volume of $<1.2\times10^{6}$ pc$^3$. Any such warm gas is likely to be confined to filament structures rather than extending throughout the cool core. 
Our non-detection of the $10^{4-5}$\,K gas suggests that the steady radiative cooling is suppressed, although the substantial systematic uncertainty does not allow us to firmly rule out the cooling flow model. 

\section{\bf Acknowledgments}

This work was supported by {\sl Chandra} Awards GO1-12160X and GO2-13125X issued by the
{\sl Chandra} X-ray Observatory Center which is operated by the Smithsonian Astrophysical Observatory under NASA contract NAS8-03060. 
The scientific results presented in this article are based on data obtained with the {\sl Hisaki} EUV observatory and {\sl Chandra} X-ray observatory. We thank Mario Chatzikos for helpful discussions. 
The authors thank the {\sl Hisaki} operation team supported by JAXA. This research has made use of the SZ-Cluster Database operated by the Integrated Data and Operation Center (IDOC) at the Institut d'Astrophysique Spatiale (IAS) under contract with CNES and CNRS.


\begin{references}
\reference{} Anderson, M. \& Sunyaev, R. 2018, A\&A, 617, 123
\reference{} Asplund, M., Grevesse, N., Jacques S.\ A. 2006, NuPhA, 777, 1
\reference{} Begelman, M.\ C. \& Fabian, A.\ C. 1990, MNRAS, 244, 26
\reference{} Bregman, J.\ N., Fabian, A.\ C., Miller, E.\ D, et al.\ 2006, ApJ, 642, 759 
\reference{} Chiang, Y.-K., Overzier, R., \& Gebhardt, K. 2013, ApJ, 779, 127
\reference{} Conselice C.\ J. \& Gallagher III J.\ S. 1999, AJ 117, 75
\reference{} Conselice C.\ J., Gallagher III J. S., Wyse R. F. G., 2001, ApJ,
122, 2
\reference{} Dere, K.\ P., Landi, E., Mason, H.\ E., Monsignori Fossi, B.\ C., \& Young, P.\ R.\ 1997, A\&AS, 125, 149
\reference{} Edgar, R.\ J. \& Chevalier, R.\ A. 1986, ApJ, 310, 27
\reference{} Edge, A.\ C.\ 2001, MNRAS, 328, 762
\reference{} Fabian, A. C.; Sanders, J. S.; Ettori, S. et al.\ 2000, MNRAS, 318, 65
\reference{} Fabian, A.\ C., Mushotzky, R.\ F., Nulsen, P.\ E.\ J. et al.\ 2001, MNRAS, 321, 20L
\reference{} Ferland, G. J., Chatzikos, M., Guzm$\rm \acute{a}$n, F. et al.\ 2017 Rev. Mex. Astron. Astrofis., 53, 385
\reference{} Forman, W. \& Jones, C. 1982, ARA\&A, 20, 547
\reference{} Forman W., Kellogg E., Gursky, H. et al.\ 1972, ApJ, 178, 309
\reference{} Hitomi Collaboration, Aharonian, F., Akamatsu, H., et al.\ 2016, Natur, 535, 117 
\reference{} Hogan, M. T., McNamara, B. R., Pulido, F. A. et al. 2017 ApJ 851, 66
\reference{} Gaspari, M., Ruszkowski, M., \& Oh, S.\ P. 2013, MNRAS, 432, 3401
\reference{} Gendron-Marsolais, M., Hlavacek-Larrondo, J., Martin, T. B., et
al.\ 2018, MNRAS, 479, L28
\reference{} Gilbank, D.\ G., Gladders, M.\ D., Yee, H.\ K.\ C. et al.\ 2011 AJ, 141, 94

\reference{} Golding, T.\ P., Leenaarts, J., Carlsson, M. 2017, A\&A, 597, 102  
\reference{} Kalberla, P.\ M.\ W., Burton, W.\ B., Hartmann, D. et al.\ 2005 A\&A 440 775
\reference{} Kuwabara, M., Yoshioka, K., Murakami, G., et al.\ 2017, JGRA, 122, 1269
\reference{} Kimura, T., Yamazaki, A., Yoshioka, K. et al.\ 2019 JSWSC, 9A 8
\reference{} Labrosse, N., Gouttebroze, P., Vial, J. -C. 2007, A\&A, 463, 117,
\reference{} Landi, E., Young, P.\ R., Dere, K.\ P.\ et al.\ 2013, ApJ, 763, 86
\reference{} McDonald, M., Veilleux, S., Rupke, D.\ S.\ N. et al.\ 2010, ApJ, 721, 2
\reference{} McCourt, M., Sharma, P., \& Quataert, E., et al.\ 2012, MNRAS, 419, 3319
\reference{} McQuinn, M. \& Switzer, E.\ R. 2010, MNRAS, 408, 1945
\reference{} McNamara, B.\ R. \& Nulsen, P.\ E.\ J. 2007, ARA\&A, 45, 117
\reference{} McNamara, B. R., Russell, H. R., Nulsen, P. E. J. et al. 2016 ApJ 830, 79
\reference{} O'Dea, C.\ P., Baum, S.\ A., Privon, G et al.\ 2008, ApJ, 681, 1035
\reference{} Oegerle, W.\ R., Cowie, L., Davidsen, A. et al.\ 2001, ApJ, 560, 187
\reference{} Panagoulia, E.\ K., Fabian, A.\ C., \& Sanders, J.\ S. 2014, MNRAS, 438, 2341
\reference{} Peterson, J.\ R., Kahn, S.\ M., Paerels, F.\ B.\ S. et al.\ 2003, ApJ, 590, 207
\reference{} Reimers, D. \& Vogel, S. 1993 A\&A 276 13 
\reference{} Russell, H.\ R., McNamara, B.\ R., Edge, A.\ C., et al.\ 2014, ApJ, 784, 78
\reference{} Russell, H.\ R., McNamara, B.\ R., Fabian, A.\ C., et al.\ 2016, MNRAS,
458, 3134
\reference{} Russell, H.\ R., McDonald, M., McNamara, B.\ R., et al.\ 2017, ApJ, 836, 130
\reference{} Salom$\rm \acute{e}$, P., Combes, F., Edge, A.\ C. et al.\ 2006, A\&A, 454, 437
\reference{} Sharon, K., Gladders, M.\ D., Marrone, D.\ P. et al.\ 2015, ApJ, 814, 21
\reference{} Su, Y., Kraft, R.\ P., Roediger, E. et al.\ 2017a, ApJ, 834, 74
\reference{} Su, Y., Nulsen, P.\.E.\ J.,  Kraft, R.\ P., et al.\ 2017b, ApJ, 847, 94 
\reference{} Sunyaev, R. A \& Zel'dovich, Y. B. 1972 CoASP 4 173
\reference{} Vantyghem, A.\ N., McNamara, B.\ R., Russell, H.\ R., et al.\ 2016, ApJ, 832, 148
\reference{} Wang, T., Elbaz, D., Daddi, E. et al.\ 2016, ApJ, 828, 56
\reference{} Werner, N., McNamara, B.\ R., Churazov, E. et al.\ 2019, SSRv, 215, 5
\reference{} Yoshioka, K., Murakami, G., Yamazaki, A. et al.\ 2013, P\&SS, 85, 250
\reference{} Yoshikawa, I., Yoshioka, K., Murakami, G. et al.\ 2014, SSRv, 184, 237
\end{references}
\end{document}